\documentclass[aps,prl,reprint,showpacs,superscriptaddress,10pt]{revtex4-1}
\usepackage{amsmath,amssymb,amsfonts,graphicx,bm,dcolumn,color}
\usepackage{mathrsfs}
\usepackage[utf8]{inputenc}
\usepackage{textcomp}
\usepackage[T1]{fontenc}
\usepackage{lmodern}

\usepackage{ulem}

\begin{document}

\title{Thermodynamics of nonadditive systems}

\author{Ivan Latella}
\email{ilatella@ffn.ub.edu}
\affiliation{Departament de F\'{i}sica Fonamental, Facultat de F\'{i}sica, Universitat de Barcelona,
Mart\'{i} i Franqu\`{e}s 1, 08028 Barcelona, Spain}

\author{Agust\'in P\'erez-Madrid}
\affiliation{Departament de F\'{i}sica Fonamental, Facultat de F\'{i}sica, Universitat de Barcelona,
Mart\'{i} i Franqu\`{e}s 1, 08028 Barcelona, Spain}

\author{Alessandro Campa}
\affiliation{Complex Systems and Theoretical Physics Unit, Health and Technology Department, 
Istituto Superiore di Sanit\`{a}, and INFN Roma 1, Gruppo Collegato Sanit\`{a}, Viale Regina Elena 299, 00161 Roma, Italy}

\author{Lapo Casetti}
\affiliation{Dipartimento di Fisica e Astronomia and CSDC, Universit\`a di Firenze,\\and INFN, Sezione di Firenze,
via G.\ Sansone 1, 50019 Sesto Fiorentino (FI), Italy}
\affiliation{INAF-Osservatorio Astrofisico di Arcetri, Largo E. Fermi 5, 50125 Firenze, Italy}

\author{Stefano Ruffo}
\affiliation{Dipartimento di Fisica e Astronomia and CSDC, Universit\`a di Firenze,\\and INFN, Sezione di Firenze,
via G.\ Sansone 1, 50019 Sesto Fiorentino (FI), Italy}

\begin{abstract}
The usual formulation of thermodynamics is based on the additivity of macroscopic systems. However, there are
numerous examples of macroscopic systems that are not additive, due to the long-range character of the interaction
among the constituents. We present here an approach in which nonadditive systems can be described within a purely
thermodynamics formalism. The basic concept is to consider a large ensemble of replicas
of the system where the standard formulation of thermodynamics can be naturally applied and the properties of a
single system can be consequently inferred. After presenting the approach, we show its implementation
in systems where the interaction decays as $1/r^\alpha$ in the interparticle distance $r$, with $\alpha$ smaller than the
embedding dimension $d$, and in the Thirring model for gravitational systems.
\end{abstract}


\newcommand{\dif}{\mathop{}\!\text{d}}
\newcommand{\ee}{\text{e}}
\newcommand{\ii}{\text{i}}
\newcommand{\kB}{k_\text{B}}
\newcommand{\sub}[1]{\text{#1}}
\newcommand{\vect}[1]{\bm{#1}}

\maketitle

Additivity plays a central role in the formulation of thermodynamics.
If a system is divided into different parts, each part possessing a certain energy, the system is said to be
additive if the energy due to the interactions between these parts is negligible in comparison with the total
energy~\cite{Campa:2009}. Due to additivity, the extensive quantities are linear functions of the system size
and the thermodynamic potentials present always the same concavity. Macroscopic systems with short-range interactions
are additive. In contrast, the energy due to interactions between different parts of the system cannot
be neglected if these interactions are long-ranged, causing the system to be intrinsically nonadditive. This lack
of additivity has been identified as the source of the unusual thermodynamic properties of systems with long-range
interactions, typically associated with a curvature anomaly of the relevant thermodynamic potential. 
The same happens in small systems with short-range interactions, where the range of the interaction is of the order
of system size~\cite{Gross:2001,Chomaz:2002}. A peculiar thermodynamic property such as negative heat
capacity is seen as unusual if one takes the thermodynamics of additive systems as a paradigm. In practice, properties
of this kind are found in a large variety of systems in nature, ranging from atomic~\cite{Schmidt:2001} to stellar
clusters~\cite{Lynden-Bell:1968,Thirring:1970,Padmanabhan:1990,Chavanis:2002:a}, so that they have become just
other common properties to be considered.

Although intense research on systems with long-range interactions has been carried out during the last years from
the statistical mechanics point of view, the thermodynamic framework concerning these systems in connection with
nonadditivity has received much less attention. This is mainly due to the fact that statistical mechanics has to
be necessarily contemplated in order to account for the microscopic interactions. Can nonadditivity be 
explicitly identified within the thermodynamic formalism, or one merely has to settle for consider it through 
its implicit contribution to the usual thermodynamic potentials? As we shall see, the thermodynamic formalism 
described here shows the clear role played by nonadditivity, which can be unambiguously determined and quantified.

The key idea is to convert the problem of the thermodynamics of a nonadditive system to the one of the thermodynamics
of an additive system and there use the standard thermodynamic approach. This can be done by considering an ensemble
of $\mathscr{N}$ independent, equivalent, distinguishable systems.
This large ensemble of replicas of the system
not necessarily has to be interpreted as a real physical system; it can be seen as a contrived system that helps
to infer thermodynamic properties of its single constituents. Since the ensemble can be as large as needed by taking
$\mathscr{N}\to\infty$, it is in fact an additive system and, therefore, the standard equilibrium thermodynamic approach
can be applied. An analogous theoretical framework was introduced by Hill for small systems~\cite{Hill:1963}.
A small system, i.e., a system with a small number of particles, is not additive; but additivity is recovered,
together with the usual thermodynamics of macroscopic systems, when the number of particles in
the system goes to infinity, provided the range of the short-range interaction becomes negligible with respect to the system size.
However, the situation we consider here is clearly different: we take from the beginning into account long-range 
interacting systems with a large number of particles. In such systems, no matter how large, the size of the system
and the interaction range are comparable, and therefore these systems are always intrinsically nonadditive.

Let us thus consider a system with energy $E$, entropy $S$, volume $V$, and $N$ particles.
We now introduce an ensemble of noninteracting replicas of the systems as a construction from which, as we will see, properties of the system itself
can be inferred. We stress that, as usual in statistical ensembles, the replicas do not interact with each other.
The total energy, entropy, volume, and number
of particles of an ensemble of $\mathscr{N}$ such systems are given by $E_\text{t}=\mathscr{N}E$,
$S_\text{t}=\mathscr{N}S$, $V_\text{t}=\mathscr{N}V$, and $N_{\text{t}}=\mathscr{N}N$, respectively. The fundamental
thermodynamic relation for the ensemble takes the form
\begin{equation}
\dif E_\text{t}=T\dif S_\text{t}-P\dif V_\text{t}+\mu\dif N_{\text{t}}+ \mathscr{E}\dif \mathscr{N},
\label{Hill_equation}
\end{equation}
where $T$ is the temperature, $P$ is the pressure exerted on the boundary of the systems,
and $\mu$ is the chemical potential of a single system. The last term on the r.h.s. of Eq.~(\ref{Hill_equation}) is the central ingredient that
this approach incorporates, which accounts for the energy variation when the
number of members of the ensemble $\mathscr{N}$ varies at constant $S_\text{t}$, $V_\text{t}$ and $N_{\text{t}}$.
The function $\mathscr{E}$ is called the replica energy and quantifies the nonadditivity of the single systems; it vanishes
for additive systems.
To see this, consider the following situation. 
Let us make a transformation in which an ensemble of $\mathscr{N}_1$ replicas, each one with $N_1$ particles, entropy $S_1$ and volume $V_1$,
becomes an ensemble of $\mathscr{N}_2$ replicas, each one with $N_2$ particles, entropy $S_2$ and volume $V_2$, under the assumption that,
for a given positive $\xi$, we have $N_2=N_1/\xi$, $S_2=S_1/\xi$, $V_2=V_1/\xi$, but $\mathscr{N}_2=\xi \mathscr{N}_1$. Clearly, in this case
$\dif S_\text{t} = \dif V_\text{t} = \dif N_\text{t} =0$, so that, from Eq.~(\ref{Hill_equation}), $\dif E_\text{t}= \mathscr{E}\dif \mathscr{N}$.
But in an additive system the energy is a linear homogeneous function of the entropy, volume and number of particles, i.e.
$E_2\equiv E(S_2,V_2,N_2)= E(S_1/\xi,V_1/\xi,N_1/\xi) = E(S_1,V_1,N_1)/\xi \equiv E_1/\xi$, and therefore $\dif E_\text{t} = 0$, requiring
$\mathscr{E}=0$. Thus, we see that additivity implies $\mathscr{E}=0$. 
Hence, $\mathscr{E}\neq0$ implies nonadditivity.
On the other hand, for a nonadditive system the energy
is not a linear homogeneous function of the entropy, volume and number of particles, and in general we will have $\mathscr{E} \neq 0$.
Thinking for example to the case $\xi=2$, this is a direct consequence of the fact that
in a nonadditive system the interaction energy between the two halves of a macroscopic system is not negligible.
Below we will show that $\mathscr{E}$ is indeed a property of the system under consideration.

Before proceeding, it is important to stress the following point. We are building a purely thermodynamic characterization
of nonadditive systems, and we have singled out one thermodynamic quantity that is peculiar for this class of systems.
However, we must be aware that one of the most striking facts in the statistical mechanics study of long-range systems,
i.e., ensemble inequivalence, should produce a correspondence in a thermodynamic treatment, since ensemble inequivalence
is connected to differences in the macroscopic states accessible to the systems when they are isolated or in contact with
a thermostat \cite{Campa:2009}. The difference in the accessible macrostates translates into a difference in the equation
of state between an isolated system and a thermostatted one, since, e.g., the temperature-energy relation of an isolated
system in the range of convex entropy cannot hold for a thermalized system~\cite{Campa:2009}. These caveats do not spoil
the central role of Eq.~(\ref{Hill_equation}) in the present treatment; one should only consider that all the thermodynamic
quantities in the equations, like e.g.\ $T$, $S$, and $\mathscr{E}$ itself, are those corresponding to the actual physical
conditions, and can be different according to whether the system is isolated or thermostatted.

Equation~(\ref{Hill_equation}) can be integrated holding all single system properties constant, 
\begin{equation}
E\dif \mathscr{N}=TS\dif \mathscr{N}-PV\dif \mathscr{N}+\mu N\dif \mathscr{N}+ \mathscr{E}\dif \mathscr{N},
\end{equation}
yielding
$E_\text{t}=TS_\text{t}-P V_\text{t}+\mu N_{\text{t}}+ \mathscr{E}\mathscr{N}$.
Thus, for a single system one has $E=TS-P V+\mu N+ \mathscr{E}$, together with the differential relations
\begin{eqnarray}
\dif S=\frac{1}{T}\dif E+\frac{P}{T}\dif V-\frac{\mu}{T}\dif N ,\\
\dif \mathscr{E}=-S\dif T+V\dif P-N\dif \mu.
\label{generalized-GD}
\end{eqnarray}
The first is the usual first law of thermodynamics; the second follows by requiring that the differentiation of
$E=TS-P V+\mu N+ \mathscr{E}$ produces the first equation. 
Moreover, Eq.~(\ref{generalized-GD}) shows
that the well-known Gibbs-Duhem equation for additive systems does not hold here, and $T$, $P$, and
$\mu$ may become independent due to the extra degree of freedom here represented by $\mathscr{E}$.
In the context of small systems, this independence between $T$, $P$, and $\mu$ has been exploited to consider completely
open liquid-like clusters in a metastable supersaturated gas phase~\cite{Hill:1998}. In addition, the deviations of
small systems thermodynamics with respect to that of macroscopic systems have been shown
in~\cite{Schnell:2011,Schnell:2012} using the grandcanonical ensemble. Furthermore, this approach has
also been used to study the critical behavior of ferromagnets~\cite{Chamberlin:2000} by considering an ensemble of
physical subdivisions of a macroscopic sample; here we always consider replicas of the whole system under consideration.

In practice, to obtain $\mathscr{E}$ one computes the entropy $S$ of a single system, whence computes the equations of
state, and thus obtains
\begin{equation}
\mathscr{E}=E-TS+PV-\mu N.
\label{replica_energy}
\end{equation}
Furthermore, to compute the entropy in a concrete case we will employ the microcanonical partition function.
Based on the considerations made before, we will then describe the thermodynamics of a long-range isolated system.

The microcanonical entropy of a single system can be obtained from phase space considerations for the whole ensemble.
Henceforth, the dimension of the embedding space is assumed to be $d$, and the position and momentum of the particle $j$
in the system $k$ are denoted by $\vect{p}_{jk}\in\mathbb{R}^d$ and $\vect{q}_{jk}\in\mathbb{R}^d$, respectively. Taking
into account that the $\mathscr{N}$ systems are independent and equivalent, the Hamiltonian $\mathcal{H}_\text{t}$ of
the ensemble is given by $\mathcal{H}_\text{t}=\sum_{k=1}^{\mathscr{N}}\mathcal{H}_k$, where each individual Hamiltonian
reads $\mathcal{H}_k=\sum_{j=1}^N|\vect{p}_{jk}|^2/(2m)+W(\vect{q}_{1k},\dots,\vect{q}_{Nk})$ with $m$ the mass of the particles.
Here $W$ is the potential
energy of a single system that contains the long-range interactions. In addition, using units where $\kB=1$, the total
entropy is given by $S_\text{t}=\ln\omega_\text{t}$, where
$\omega_\text{t}=\omega_\text{t}(E_\text{t},V_\text{t},N_\text{t},\mathscr{N})$ is the density of states obtained from
the phase space of the ensemble. This density of states must be computed not only considering that the total energy is
fixed to $E_\text{t}=\mathscr{N}E$, but also that the energy of each single system is fixed to $E$. According to this,
the energy constraint in phase space can be written as
$\rho(E_\text{t})= \prod_{k=1}^\mathscr{N}\delta(E_\text{t}/\mathscr{N}-\mathcal{H}_k)$. Since the systems are also
considered to be distinguishable, the microcanonical density of states of the ensemble is therefore given by
\begin{equation}
\omega_\text{t}= \int \frac{\rho(E_\text{t})}{\left(h^{dN}N!\right)^{\mathscr{N}}}\prod_{k=1}^\mathscr{N}\dif^{2dN}\vect{\Gamma}_k
=\omega^\mathscr{N},
\label{total_density_of_states}
\end{equation}
where $h$ is a constant, $\dif^{2dN}\vect{\Gamma}_k=\prod_{j=1}^N\dif^d\vect{q}_{jk} \dif^d\vect{p}_{jk}$, and
$\omega=\omega(E,V,N)$ is the density of states of a single system. Since the particles are confined to move within
the walls of their own system, spatial integrations in (\ref{total_density_of_states}) extend over the volumes $V_k$
satisfying $\int_{V_k}\dif^d\vect{q}_{jk}=V$ for all $j$ and $k$. Thus, in view of (\ref{total_density_of_states}),
as required, $S_\text{t}=\mathscr{N}\ln\omega=\mathscr{N}S$, which highlights the fact that the information concerning
nonadditivity is contained in the microcanonical entropy $S$ of a single system.

The contribution of long-range interactions to (\ref{replica_energy}) can be further concretized if the thermodynamic
quantities are separated into a part evaluated without long-range interactions and the corresponding excess produced by
these interactions. Assuming that there are no short-range interactions, so that the potential energy is only due to
long-range interactions, we write the entropy and energy as $S=S^{(\text{i})}+S^{(\text{e})}$ and
$E=E^{(\text{i})}+E^{(\text{e})}$, respectively, with $E^{(\text{e})}=W$. Here, quantities labeled with (i) correspond 
to the ideal gas contribution, while (e) indicates the excess produced by the long-range interactions.
Analogously, we write $\mu=\mu^{(\text{i})}+\mu^{(\text{e})}$ and
$P=P^{(\text{i})}+P^{(\text{e})}$. Using these expressions, from (\ref{replica_energy}) one obtains
\begin{equation}
\mathscr{E}=W-TS^{(\text{e})}+P^{(\text{e})}V-\mu^{(\text{e})} N
\label{replica_energy_excess}
\end{equation}
since in the absence of long-range interactions $TS^{(\text{i})}=E^{(\text{i})}+P^{(\text{i})}V-\mu^{(\text{i})} N$.
If we include also short-range interactions, and not only ideal contributions, the last statement is still true
if the splitting to account for the excess produced by long-range interactions is performed in such a way that Eq.~(\ref{replica_energy_excess}) is satisfied.
Expression (\ref{replica_energy_excess}) for the replica energy
can be significantly simplified using the mean-field approximation in the large $N$ limit; an approximation
that can be employed for long-range systems~\cite{Mori:2013}. This is discussed next.

The mean-field potential energy of a single system can be written as
$W=\frac{1}{2}\int n(\vect{x})\Phi(\vect{x})\dif^d \vect{x}$, where $n(\vect{x})$ is the number density and
$\Phi(\vect{x})$ is the potential characterizing the long-range interactions at a point $\vect{x}\in\mathbb{R}^d$ in
the one-particle configuration space of a single system. Likewise, the mean-field entropy of a single system takes the form
\begin{equation}
S=-\int n(\vect{x})\ln\left[\lambda_T^d n(\vect{x})\right]\dif^d \vect{x}+\frac{2+d}{2}N,
\label{entropy}
\end{equation}
where $\lambda_T=h/\sqrt{2\pi mT}$ is the thermal wavelength. Since long-range interactions act locally as an external field, the
global chemical potential takes the form~\cite{Latella:2013} 
\begin{equation}
\mu=\mu_0(\vect{x})+\Phi(\vect{x}), 
\label{mu}
\end{equation}
which, being constant, prevents net fluxes of particles through the system. Here $\mu_0(\vect{x})$ is the local ideal
chemical potential, which depends on $n(\vect{x})$ in such a way that from~(\ref{mu}) one obtains
$n(\vect{x})= \lambda_T^{-d} \exp\left\{-[\Phi(\vect{x})-\mu]/T\right\}$. Furthermore, multiplying both sides
of (\ref{mu}) by the number density and integrating over the volume one obtains
\begin{equation}
\mu N=T\int\dif^d \vect{x}\ n(\vect{x})\ln\left[\lambda_T^d n(\vect{x})\right]+2W.
\label{mu_N}
\end{equation}

Once suitable expressions for the entropy and chemical potential have been derived, the relation between their
excess parts can be explicitly written down. Since $S^{(\text{i})}$ and $\mu^{(\text{i})}$ can be obtained
from (\ref{entropy}) and (\ref{mu_N}), respectively, by setting $\Phi(\vect{x})=0$, the excess quantities
$S^{(\text{e})}=S-S^{(\text{i})}$ and $\mu^{(\text{e})}=\mu-\mu^{(\text{i})}$ follow straightforwardly. As a
consequence, one has $\mu^{(\text{e})} N=-TS^{(\text{e})}+2W$, and therefore
\begin{equation}
\mathscr{E}=-W+P^{(\text{e})}V.
\label{replica_energy_excess_reduced}
\end{equation}
Equation (\ref{replica_energy_excess_reduced}) for the replica energy
is particularly useful since it does not involve entropy and chemical potential.

As examples of systems where
$\mathscr{E}$ can be computed exactly using the mean-field description, we have systems where the interactions decay
as $1/r^\alpha$ in the interparticle distance $r$, with $0\leq\alpha\leq d$. In this case, the virial theorem states
that $dNT+\alpha W=dPV$ and hence $P^{(\text{e})}V=\alpha W/d$. Thus, the
replica energy becomes~\cite{Latella:2013} $\mathscr{E}=-(1-\alpha/d) W$,
and we see that the system becomes additive for the marginal case $\alpha=d$. In the case of infinite-range, nondecaying
interactions, i.e., $\alpha=0$, the system is homogeneous and thus one has
$TS=E^{(\text{i})}+P^{(\text{i})}V-\mu^{(\text{i})} N$. This apparent inconsistency is solved by realizing that
$\alpha=0$ means actually no interaction at all, since $W$ is a constant; therefore the vanishing of the potential
at infinite distance requires $W\equiv 0$, and then $\mathscr{E}=0$. Also, note that for $\alpha>d$ the mean-field approximation
fails, and therefore one cannot use the last expression to infer that $\mathscr{E}\ne 0$.

\begin{figure}
\includegraphics{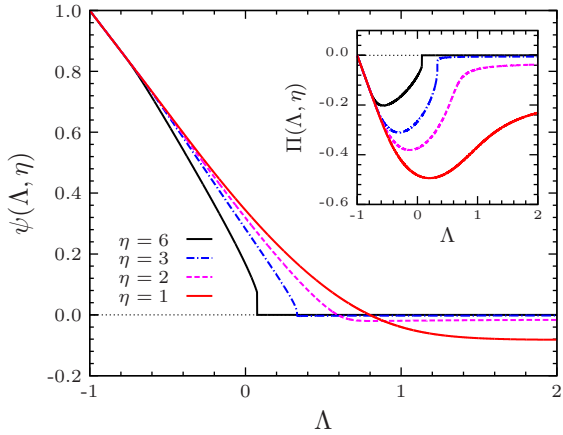}
\caption{Microcanonical reduced replica energy for the Thirring model
as a function of the reduced energy for different values of $\eta$. In the inset we show
$\Pi(\Lambda,\eta)=P^{(\text{e})}V/(\nu N^2)$, which is one of the contributions to the reduced
replica energy; the other contribution, $-W/(\nu N^2)$, decreases for
increasing $\Lambda$ and $\eta$.}
\label{Figure:1}
\end{figure}

Another example that we will consider is the Thirring model~\cite{Thirring:1970} in $d=3$.
This is a solvable model that incorporates some of the remarkable features of systems with long-range interactions.
In this model, inside the volume $V$ of the system there is a core of volume $V_\text{c}$ where all particles in its
interior interact uniformly with each other. Outside the core, the particles do not interact and thus behave as a
free gas. For large $N$, the mean field potential can be written as
$\Phi(\vect{x})=-2\nu N_\sub{c} \theta_{V_\sub{c}}(\vect{x})$, where $\nu$ is a constant, $N_\sub{c}$ is the number
of particles in $V_\sub{c}$, and $ \theta_{V_\sub{c}}(\vect{x})=1$ if $\vect{x}\in V_\sub{c}$ and vanishes otherwise.
Thus, introducing the number of free-gas particles $N_\sub{g}=N-N_\sub{c}$, the potential energy of the system reads
$W= -\nu(N-N_\sub{g})^2$.  Let us also introduce the reduced variables $\Lambda\equiv E/(\nu N^2)$ and
$\eta\equiv\ln(V/V_\sub{c}-1)$. For fixed $E$, $V$, and $N$, and according to the saddle-point method, the fraction
$\bar{n}_\sub{g}(\Lambda,\eta)=N_\sub{g}/N$ that defines the equilibrium states of the system is obtained by
solving~\cite{Thirring:1970}
\begin{equation}
\frac{3\left(1-\bar{n}_\text{g}\right)}
{\Lambda+\left(1-\bar{n}_\text{g}\right)^2}-\ln\left(\frac{1-\bar{n}_\text{g}}{\bar{n}_\text{g}}\right)-\eta=0.
\label{condition_maximum}
\end{equation}
The microcanonical entropy per particle $s(\Lambda,\eta)\equiv S(E,V,N)/N$ takes the form
\begin{eqnarray}
s(\Lambda,\eta)&=& \frac{3}{2}\ln\left[\Lambda+\left(1-\bar{n}_\text{g}\right)^2\right]-
(1-\bar{n}_\text{g})\ln(1-\bar{n}_\text{g})\nonumber\\
&&-\bar{n}_\text{g} \ln \bar{n}_\text{g}+\bar{n}_\text{g}\eta-\ln\left(\ee^\eta+1\right)+s_0 \\
&&\equiv s_1(\Lambda,\eta) + s_0 \nonumber,
\label{entropy_per_particle}
\end{eqnarray}
with $s_0=\ln\left(cVN^{1/2}\right)$, being $c$ a constant.
Accordingly, the temperature reads as $T=\frac{2}{3}\nu N \left[\Lambda+\left(1-\bar{n}_\text{g}\right)^2\right]$.

Since the particles outside the core are free, the pressure $P$ at the boundary of the system is clearly given
by $P(V-V_\sub{c})=N\bar{n}_\sub{g}T$. Thus, from (\ref{replica_energy_excess_reduced}) one has
\begin{equation}
\psi\equiv\frac{\mathscr{E}}{\nu N^2}= \left(1-\bar{n}_\text{g}\right)^2+\tau\left[\bar{n}_\text{g}\left(1+\ee^{-\eta}\right)-1\right],
\label{psi}
\end{equation}
where $\tau\equiv T/(\nu N)$.
The function $\psi$ is the reduced microcanonical replica energy.
Using (\ref{condition_maximum}) to express $\ee^{-\eta}$ in (\ref{psi}), it is not difficult to see that for
$\bar{n}_\text{g}>0$ and $\Lambda\gg 3\left(1-\bar{n}_\text{g}\right)$,
$\psi$ approaches $W/(\nu N^2)=-\left(1-\bar{n}_\text{g}\right)^2$,
which, in turn, decreases in modulus for increasing $\eta$. This behavior can be seen in Fig.~\ref{Figure:1} for different
values of the reduced volume $\eta$. Notice that in the plot $\psi$ presents a jump at a value of $\Lambda$ slightly
greater than zero for $\eta=6$; this is because the model possesses a first-order microcanonical phase transition and the
replica energy contains this information. 

\begin{figure}
\includegraphics{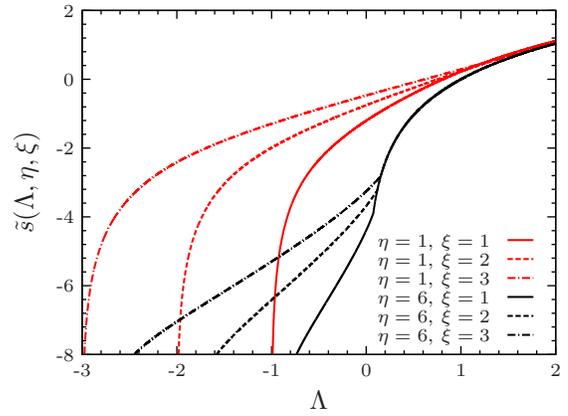}
\caption{Scaled microcanonical entropy per particle for the Thirring model as a function of $\Lambda$ for different
values of the scale factor and the reduced volume $\eta$. The curves get close to each other for values of $\Lambda$
for which the replica energy becomes relatively small.
Even for high energies, nonadditivity makes impossible a strict scaling regime.}
\label{Figure:2}
\end{figure}

Furthermore, in order to evince the intrinsic nonadditivity of the system, it is instructive to study how the entropy
behaves when the thermodynamic variables are scaled. With this purpose, we introduce a scale factor $\xi\geq1$ and a
scale transformation that acts on a quantity $Q$ such that $Q'=\xi Q$. Since $E'/(\nu N'^2)=\Lambda/\xi$ and $V'=\xi V$,
the dimensionless parameters $\Lambda$ and $\eta$ transform according to $\Lambda\to\tilde{\Lambda}=\Lambda/\xi$ and
$\eta\to\tilde{\eta}=\ln\left[\xi\left(\ee^\eta+1\right)-1\right]$. We can write
$S'(E',V',N')/N'= s_1(\tilde{\Lambda},\tilde{\eta})+ \ln\left(cVN^{1/2}\right)+\frac{3}{2}\ln\xi$, where the last
term comes from the scaling of the volume and the number of particles in $s_0$. The fraction $\bar{n}_\sub{g}$ now is
obtained from~(\ref{condition_maximum}) with $\tilde{\Lambda}$ and $\tilde{\eta}$ in the place of $\Lambda$ and $\eta$,
respectively. Moreover, the entropy per particle of the fully scaled system can be expressed as a function of the parameters
of the system without scaling by defining $\tilde{s}(\Lambda,\eta,\xi)\equiv s_1(\tilde{\Lambda},\tilde{\eta})+\frac{3}{2}\ln\xi$,
where we have subtracted $s_0$ for convenience since it does not depend on $\xi$. The scaled entropy per particle $\tilde{s}$
is shown in Fig.~\ref{Figure:2}. Due to the nonadditivity, it is clearly seen that the entropy strongly depends on the scale
transformation, as expected. As the energy increases, however, in this case the system becomes, so to speak, more additive
since the curves tend to run together, although they never touch each other. If the entropy were a linear homogeneous function
of $E$, $V$, and $N$, all curves would collapse into a single one. The interesting fact is that the nonadditivity becomes
less noticeable when the replica energy is relatively small, as should be
expected since $\mathscr{E}=0$ for additive systems.

While in a statistical mechanics formulation the nonadditivity is naturally codified in the
Boltzmann-Gibbs microcanonical entropy, as well as in the corresponding free energy for other ensembles,
we have seen here that, in the thermodynamic treatment, nonadditivity emerges through an additional degree of
freedom, the replica energy $\mathscr{E}$. According to the differences between
isolated systems and systems in contact with a thermostat emphasized before, we expect that the
replica energy will depend on the physical situation under consideration.
Nevertheless, in a long-range systems it will always be different from zero. In conclusion, we have shown that
nonadditive systems can be treated in the standard equilibrium thermodynamic framework if it is properly formulated.

\begin{acknowledgments}
We would like to thank Dick Bedeaux, Ralph Chamberlin and \O{}ivind Wilhelmsen for fruitful and
stimulating discussions. I.L. acknowledges financial support through an FPI scholarship (Grant No. BES-2012-054782) from
the Spanish Government. This work was partially supported by the Spanish Government under Grant No. FIS2011-22603.
We also thank the Galileo Galilei Institute for Theoretical Physics for the hospitality and the INFN for partial support
during the completion of this work. 
\end{acknowledgments}

\end{document}